\pdfoutput=1

\documentclass[11pt]{article}

\usepackage[preprint]{acl}

\usepackage{multirow}
\usepackage{graphicx}
\usepackage{float}
\usepackage{colortbl}
\usepackage{subfig}
\usepackage{enumitem}
\usepackage{booktabs}
\usepackage{makecell}
\usepackage{subcaption} 
\usepackage{times}
\usepackage{latexsym}

\usepackage[T1]{fontenc}

\usepackage[utf8]{inputenc}

\usepackage{microtype}

\usepackage{inconsolata}

\usepackage{graphicx}
\usepackage{amsmath}
\newcommand{\wcb}[1]{{\color{black}{#1}}}
\newcommand{\wang}[1]{{\color{black}{#1}}}
\newcommand{\wroll}[1]{{\color{black}{#1}}}
\newcommand{\ww}[1]{{\color{black}{#1}}}
\newcommand{\www}[1]{{\color{black}{#1}}}

\newcommand{\roll}[1]{{\color{black}{#1}}}

\title{Decoding in Latent Spaces for Efficient Inference in LLM-based Recommendation}

\author{
 \textbf{Chengbing Wang\textsuperscript{1}},
 \textbf{Yang Zhang\textsuperscript{2}\thanks{Corresponding author}}, 
 \textbf{Zhicheng Wang\textsuperscript{1}},
 \textbf{Tianhao Shi\textsuperscript{1}},
\\
 \textbf{Keqin Bao\textsuperscript{1}},
 \textbf{Fuli Feng\textsuperscript{1}\footnotemark[1]},
 \textbf{Tat-Seng Chua\textsuperscript{2}},
\\
\\
 \textsuperscript{1}University of Science and Technology of China,
 \textsuperscript{2}National University of Singapore,
\\
 \texttt{\{wwq197297,wangzhicheng,sth,baokq\}@mail.ustc.edu.cn,}\\
 \texttt{\{zyang1580, fulifeng93\}@gmail.com, dcscts@nus.edu.sg}
}

\begin{document}
\maketitle
\begin{abstract}

Fine-tuning large language models (LLMs) for recommendation in a generative manner has delivered promising results, but encounters significant inference overhead due to autoregressive decoding in the language space. 
This work explores bypassing language-space decoding by directly matching candidate items with the LLM's internal thought representations in the latent space, eliminating the time-consuming autoregressive process to reduce computational costs.
Towards this, we introduce \textit{Light Latent-space Decoding} ($L2D$), an effective and efficient latent-space decoding method. 
$L2D$ represents user-preferred items by using the hidden states of test sequences reflecting the LLM’s internal thought, and obtains candidate item representations from the hidden states of training sequences labeled with the corresponding candidate items.
It then matches the two types of representations to decode items, achieving latent-space decoding. 
In this way, it enables efficient decoding without altering the LLM's generative tuning paradigm, thereby preserving performance.
Extensive empirical results demonstrate that $L2D$ is more than 10x faster than language-space decoding while maintaining or enhancing performance.


\end{abstract}

\section{Introduction}

Large language models (LLMs) have been widely fine-tuned on recommendation data in textual format to directly generate the next item of user interest based on historical interactions~\cite{Re_21,Re_23,bigrec,tallrec,zheng2024adapting}. This training paradigm aligns well with the generative nature of LLMs, enabling them to develop sophisticated user understanding and interest-mining capabilities. When deploying these fine-tuned LLMs for personalized recommendations (LLM4Rec), a key research challenge is how to effectively decode the items that LLM truly "thinks" or prefers internally.

Current LLM4Rec methods primarily rely on the LLM’s internal decoding ability (\textit{i.e.,} language-space decoding), which typically operates the LLM in an autoregressive token-by-token manner to output the item representations in the language space (e.g., title)~\cite{bigrec}. 
In the autoregressive decoding, one token generation would wait for all preceding tokens' generations, incurring substantial time costs. 
Worse yet, each recommendation request from users typically requires generating a list of items~\cite{lin2024efficient,d3,zheng2024adapting}, linearly scaling the cost regarding the list size.
While grounding techniques~\cite{bigrec} can reduce these costs by mapping each generated item to multiple actual items, they may lead to performance degradation. For example, in our findings, mapping only one generated item for top-10 recommendations would result in a performance drop of fifty percent compared to generating 10 items (\textit{c.f.,} Figure~\ref{fig:balance}).

\begin{figure}
    \centering
    \includegraphics[width=1\linewidth]{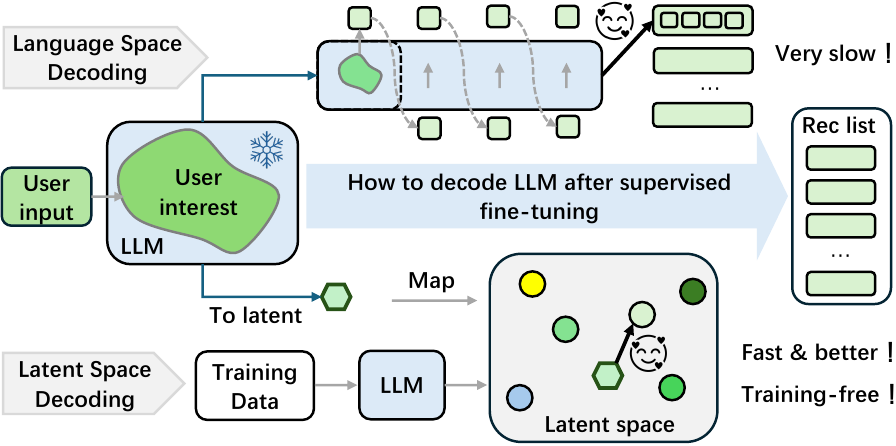}
    \caption{
    Illustration of language-space vs. latent-space decoding: Latent-space decoding bypasses the slow language-space decoding and instead achieve decoding via directly matching candidate item with the LLM internal 'thought' items in the latent space. It preserves the generative tuning paradigm to keep performance while enabling efficient decoding.
    } 
    \label{fig:intro_method}
    \vspace{-1em}
\end{figure}


This work considers bypassing this language-space decoding to enhance decoding efficiency. Training a recommendation head to decode the next item by directly predicting its ID provides a straightforward way to avoid language-space decoding, improving decoding efficiency. However, this \ww{training} objective deviates from the language model’s original goal of next-token prediction, which may hinder effective utilization of its pretrained knowledge. 
This raises a new question: can we bypass language-space decoding while still preserving the powerful generative training characteristics of the LLM's recommendation tuning?



By re-examining LLM decoding, we can interpret the tuned LLM as having already thought a recommendation target within its internal latent space in response to a user query, with autoregressive decoding merely serving to activate it. If we can find the appropriate representation for the item in this latent space—and represent all candidate items similarly—then efficient decoding can be achieved through representation matching within the same space. Meanwhile, since this approach only extracts the LLM’s internal “thoughts” learned \www{during LLM's generative training phase}, it preserves the original generative training objective during recommendation tuning.     

To obtain representations of the items internally considered by the LLM, we propose using the last hidden states from the final LLM layers corresponding to a given test sample, as the state primarily drives item generation. For candidate items, since the training set already provides matching pairs of hidden states and the item serving as ground-truth items, we can aggregate all the hidden states associated with the item to form an effective representation in the latent space, without incurring additional training cost. 
This approach is based on the idea that each paired hidden state captures a distinct feature aspect of the corresponding ground-truth item, allowing us to combine them into a meaningful and comprehensive target representation.

To this end, we propose \textit{Light Latent-space Decoding ($L2D$)}, a simple yet efficient method for latent-space decoding. After finishing generative training, we store the training samples' hidden states and their labels (\textit{i.e.}, \wang{ground-truth} items) in a memory module and create each item's representation by aggregating its associated hidden states in the memory. Then we decode items to recommend by finding the item whose representation is most similar to the test sample's hidden state using L2 distance. Regarding the aggregation to form item representation, $L2D$ offers two strategies: 1) global aggregation, which averages all associated hidden states for an item, and 2) local aggregation, which uses only the top-$M$ most similar samples from the memory based on the test sample's hidden state. The global strategy provides a comprehensive representation, while the local strategy focuses on aspects most relevant to the test sample.  

The main contributions of this work are summarized as follows:
\begin{itemize}[leftmargin=*, itemsep=0pt,parsep=1pt]
\item \ww{We propose bypassing language-space decoding for efficient recommendation inference while preserving the powerful generative training characteristics of LLM for recommendation.}
\item We introduce $L2D$, a simple yet effective method that performs latent-space decoding by leveraging the hidden states of test and training sequences, to capture the LLM’s internal thought.
\item \roll{Extensive experiments demonstrate that applying $L2D$ to existing LLM-based recommendation methods reduces inference latency by at least 10 times compared to language-space decoding while maintaining or enhancing performance.}

\end{itemize}

\section{LLM-based Generative Recommender}

Let $\mathcal{D}$ represent the user-item interaction data. The $j$-th sample in $\mathcal{D}$ is denoted as $(s_j, v_j)$, where $s_j$ represents a user's interaction history, and $v_j$ is the interacted item for the sample. Notably, both $s_j$ and $v_j$ are in textual form. To train an LLM-based generative recommender, we convert each sample $(s_j, v_j)$ into instruction data, using a fixed prompt template such as "\textit{A user has interacted with the following items: <$s_j$>; which item would the user like next?}", with $v_j$ as the ground-truth model output. Then, the instruction data $\{(prompt(s_j), v_j)\}_{\mathcal{D}}$ can be utilized to fine-tune the LLM.

During inference, given a user's interaction history $s$ to generate the next item, the LLM first encodes the prompt into hidden states, formally:
\begin{equation}\small
\label{eq:hidden-state}
    h = LLM_{last}(prompt(s)),
\end{equation}
where $h$ denotes the \textbf{last} hidden state of the input $prompt(s)$ at the final layer, and $LLM_{last}(\cdot)$ represents the function that extracts the hidden state from the last layer of the LLM. In the language-space decoding method, $h$ is further mapped to the LLM's output layer to generate the first item token, which is then added to the input, and the process repeats to generate a full item. In contrast, we explore decoding items from the hidden state $h$.


\section{Latent-Space Decoding}

In this section, we introduce our \textit{Light Latent-space Decoding ($L2D$)} framework, starting with presenting the overview and followed by a detailed description of its key components.



\begin{figure*}
    \centering
    \includegraphics[width=0.9\linewidth,height=6.2cm]{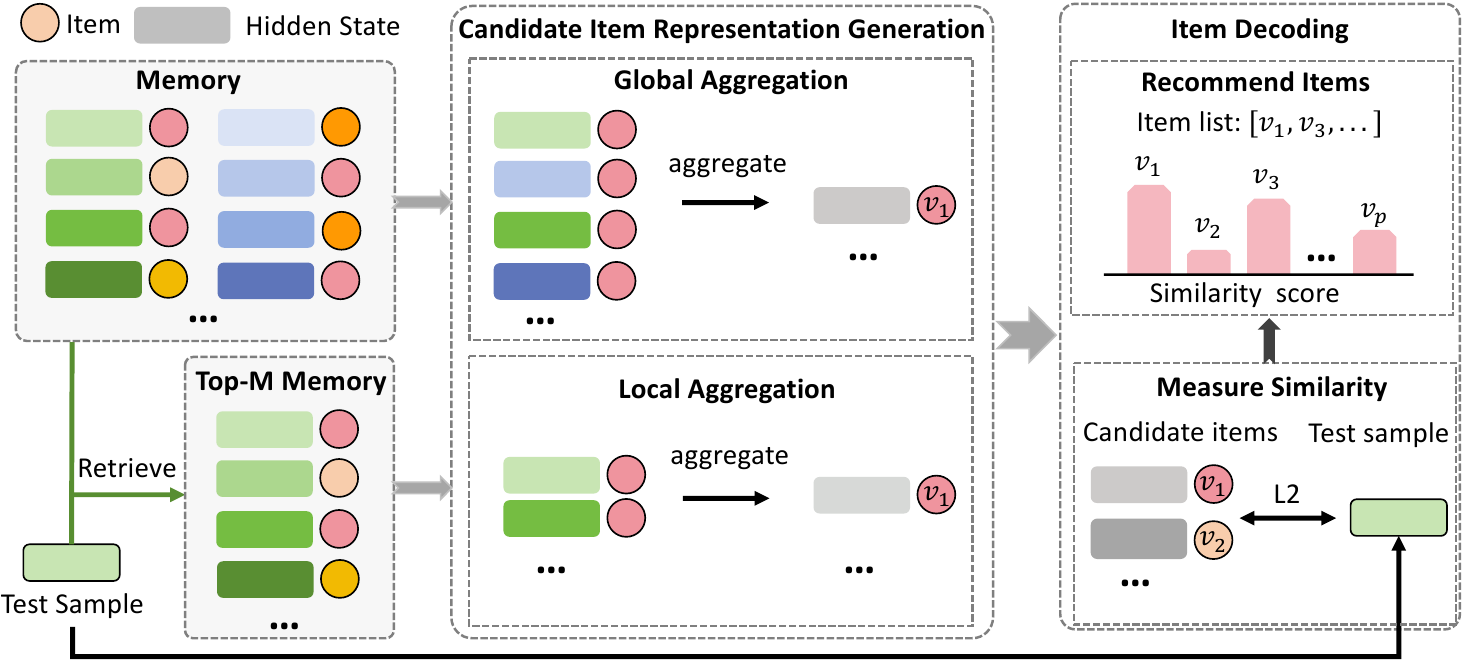}
    \caption{The overview framework of our proposed $L2D$. 
    The left part illustrates the memory set that stores (hidden state, ground-truth item) pairs. 
    The middle part illustrates how $L2D$ generates candidate item representations via global aggregation (averaging all associated hidden states) or local aggregation (using the top-M relevant samples to the test sample).
    The right part depicts the item decoding phase by measuring the similarity between the test sample's hidden state and candidate item representations.
    }
    \label{fig:framework}
    \vspace{-1em}
\end{figure*}

\subsection{Overview}





The main idea of this work is to bypass time-intensive language-space decoding while still preserving the powerful generative training characteristics of the LLM's recommendation tuning.  
To achieve this, we propose $L2D$, a light latent-space decoding framework, which directly utilizes the hidden states from the LLM to construct latent space for decoding, where the LLM is trained with recommendation data in a generative manner.
Figure~\ref{fig:framework} illustrates the overall $L2D$ process, which consists of three steps:  
\begin{itemize}[leftmargin=*, itemsep=0pt,parsep=1pt]
\item[1)] \textbf{Memory Construction}: Stores (hidden state, ground-truth item) pairs from training samples in a memory module, preparing for candidate item representation generation. 
\item[2)] \textbf{Candidate Item Representation Generation}: Produces representations for each item by aggregating its associated hidden states stored in memory. 
\item[3)] \textbf{Item Decoding}: Matches the hidden state of a test sample with the candidate item representations to determine the output. 
\end{itemize}
The first step can be pre-computed, ensuring no impact on inference latency, while the last two steps operate independently of LLM, minimizing latency. We provide detailed explanations below.





\subsection{Memory Construction}

$L2D$ begins by constructing a memory set that stores the (hidden state, ground-truth item) pairs from the training samples. 
Specifically, for the $j$-th training sample $(s_j, v_j)$, we compute its last hidden state at the final layer using Equation~\eqref{eq:hidden-state} as \( h_j = LLM_{last}(prompt(s_j)) \) and store the pair \((h_j, v_j)\) in a memory set \(\mathcal{M}\). Repeating this process for all samples in the training set, \(L2D\) constructs the final memory \(\mathcal{M}\), formally,
\begin{equation}
    \mathcal{M}=\{(h_j,v_j) \mid j=1,\dots,N\},
\end{equation}
where $N$ denotes the total number of training samples.
The hidden state in each pair reflects a specific feature aspect of the corresponding item, meaning the process effectively captures one facet of the LLM’s original understanding of the item in the same latent space. The memory set is then used to generate representations of candidate items in this latent space.

\subsection{Candidate Item Representation Generation}

After constructing the memory, \( L2D \) leverages the stored (hidden state, ground-truth item) pairs to generate representations of candidate items in the latent space. For each candidate item, it aggregates the associated hidden states—those paired with the item as the ground-truth item—to create the item's representation.
In particular, $L2D$ offers two aggregating strategies: 1) global aggregation, which averages all associated hidden states for each candidate item, and 2) local aggregation, which uses only the top-$M$ most similar samples in the memory based on the test sample's hidden state. The global strategy provides comprehensive representation of candidate items, while the local strategy makes the representation more relevant to the test samples. We will first elaborate on the two strategies, followed by a comparison.

\noindent $\bullet$ \textbf{Global Aggregation}.  
To aggregate the hidden states stored in the memory \(\mathcal{M}\) for creating representation of candidate items, a straightforward approach is to directly average all hidden states associated with the same item. The global aggregation follows this strategy.  
Specifically, we first group hidden states in memory by items and then average the hidden states within each group to form the corresponding item's representation. Formally, for an item \(v\), its representation $\bar{h}_v$ is computed as follows:  
\begin{equation}\label{eq:global-rep}
\bar{h}_v = \frac{1}{|\mathcal{M}(v)|} \sum_{h_j \in \mathcal{M}(v)} h_j, 
\end{equation}
where \(\mathcal{M}(v)\) denotes the set of all hidden states associated with item \(v\), defined as  
\[
\mathcal{M}(v) = \{h_j \mid  (h_j, v_j)\in \mathcal{M}, v_j = v\}.
\]
The size of \(\mathcal{M}(v)\) is denoted by \(|\mathcal{M}(v)|\).

\noindent$\bullet$\textbf{Local Aggregation}.
The LLM's understanding of a candidate item may encompass multiple feature aspects, and the global aggregation method combines all aspects to form a comprehensive item representation. However, during the inference stage, not all feature aspects are relevant for each test sample; only the aspects related to the test sample are important. This suggests that mixing all feature aspects in one representation may introduce interference. With this in mind, we propose local aggregation, which leverages only the top-$M$ samples from memory that are most relevant to the test sample's hidden state for item representation generation. 

Specifically, for a test sample with \(s_t\), we first filter a subset of the memory based on the hidden state \(h_t\) of test sample, denoted as \(\mathcal{M}_t\). Formally, 
\begin{equation}
\begin{aligned}
\mathcal{M}_{t} = 
&\left\{
(h_j, v_j) \mid  (h_j, v_j) \in \mathcal{M} ,  \right . \\
& \left. S(h_t, h_j) \text{ is in the top-$M$ largest} 
\right\},
\end{aligned}
\end{equation}
where \(S(h_t, h_j) = \frac{1}{\|h_t - h_j\|_2}\) measures the similarity between the stored hidden state \(h_j\) and the test sample's hidden state \(h_t\).  
Then, a process similar to global aggregation is applied to \(\mathcal{M}_t\) to obtain the candidate item representation. Given a candidate item \(v\), the representation is formulated as follows:  
\begin{equation}\label{eq:local-rep}
    \bar{h}_v^t = \frac{1}{|\mathcal{M}_t(v)|} \sum_{h_j \in \mathcal{M}_t(v)} h_j,
\end{equation}
where \(|\mathcal{M}_t(v)|\) denotes the size of \(\mathcal{M}_t(v)\), and \(\mathcal{M}_t(v)\) is the subset of \(\mathcal{M}_t\) containing items with \(v\) as the ground-truth, defined as  
\[
\mathcal{M}_t(v) = \{ h_j \mid  (h_j, v_j) \in \mathcal{M}_t, \, v_j = v \}.
\]

\noindent\textbf{Global vs. Local Aggregation:} Compared to global aggregation, local aggregation can better focus on test sample-specific aspects, potentially improving subsequent matching performance. However, it may struggle more with sparse items due to an increased lack of associated hidden states. Additionally, unlike the representation obtained through global aggregation, which is uniform for all test samples, the representation derived from local aggregation is tailored to each test sample. For quantitative comparative experiments and further analysis of their suitable scenarios for the global and local aggregation, please refer to Section~\ref{Sparse and Dense}. Meanwhile, the computational cost of both aggregation methods remains negligible, as our approach fundamentally bypasses the time-consuming autoregressive decoding of the LLM. The method requires only a single forward pass during LLM inference, with all subsequent operations being performed through efficient vector-level computations. 

\begin{figure*}[tbp]
  \centering            
  \subfloat{
\label{fig:CD_balance}\includegraphics[width=0.39\textwidth]{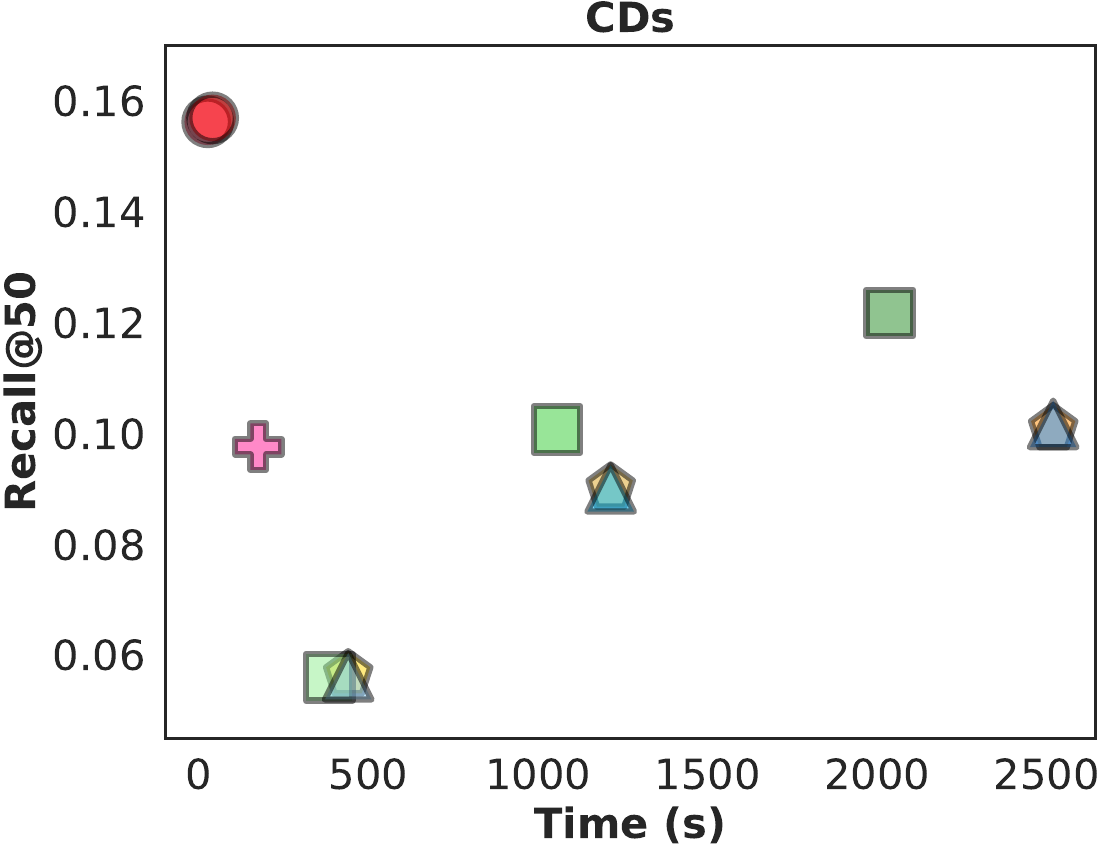}}
  \subfloat{
    \label{fig:Game_balance}\includegraphics[width=0.525\textwidth]{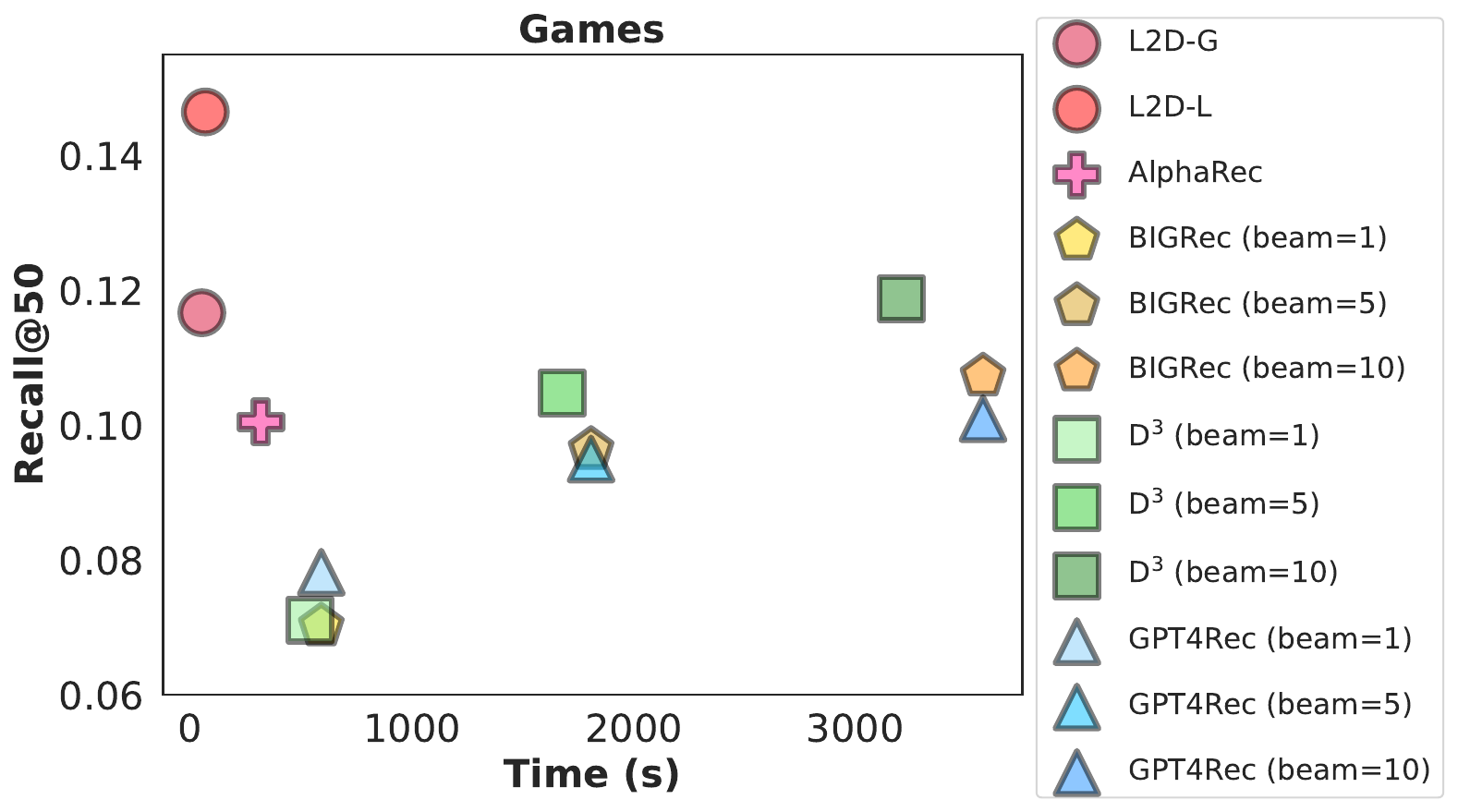}}
    \vspace{-5pt}
  \caption{The Recall@50 performance and the inference overhead of LLM-based recommenders on two datasets.
}
  \label{fig:balance}
\end{figure*}

\begin{table*}[ht]
\centering
\renewcommand{\arraystretch}{1.3}
\caption{\roll{Overall performance comparison. \textbf{Results with beam size 1 are reported for methods using beam search for fair comparison, with results for other beam sizes in Figure~\ref{fig:balance}}. The best results are in bold.}
}
\vspace{-5pt}
\resizebox{0.95\linewidth}{!}{
\begin{tabular}{l|l|llllll|llllll}
\hline
&         & \multicolumn{6}{c|}{CDs}   & \multicolumn{6}{c}{Games}   \\ \hline
&Model    & R@20            & R@50            & \multicolumn{1}{l|}{R@100}           & N@20            & N@50            & N@100           & R@20            & R@50            & \multicolumn{1}{l|}{R@100}           & N@20            & N@50            & N@100           \\ \hline
\multirow{2}{*}{Traditional}&SASRec   & 0.1015 & 0.1271 & \multicolumn{1}{l|}{0.1522} & 0.0602 & 0.0653 & 0.0693  & 0.0684 & 0.1117 & \multicolumn{1}{l|}{0.1564} & 0.0332 &0.0417 & 0.0490  \\
&GRU4Rec  & 0.0707 & 0.1027 & \multicolumn{1}{l|}{0.1347} & 0.0376 & 0.0439 & 0.0491 & 0.0664 & 0.1099 & \multicolumn{1}{l|}{0.1601} & 0.0302 & 0.0387 & 0.0468  \\ \hline

LLM embedding &AlphaRec & 0.0651          & 0.0976          & \multicolumn{1}{l|}{0.1353}          & 0.0300           & 0.0364          & 0.0425          & 0.0619          & 0.1005          & \multicolumn{1}{l|}{0.1392}          & 0.0295          & 0.0371          & 0.0434          \\ \hline
\multirow{3}{*}{\makecell{LLM Generative \\ (beam = 1)}}&GPT4Rec  & 0.0513          & 0.0562          & \multicolumn{1}{l|}{0.0652}          & 0.0433          & 0.0443          & 0.0458          & 0.0508          & 0.0782          & \multicolumn{1}{l|}{0.1064}          & 0.0293          & 0.0347          & 0.0392          \\
&BIGRec   & 0.0506          & 0.0565          & \multicolumn{1}{l|}{0.0621}          & 0.0435          & 0.0446          & 0.0456          & 0.0476          & 0.0702          & \multicolumn{1}{l|}{0.1007}          & 0.0284          & 0.0328          & 0.0378          \\
&$D^3$    & 0.0507          & 0.0560          & \multicolumn{1}{l|}{0.0623}          & 0.0436          & 0.0447          & 0.0457          & 0.0478          & 0.0711          & \multicolumn{1}{l|}{0.1004}          & 0.0284          & 0.0330          & 0.0376          \\ \hline
\multirow{2}{*}{Ours}&$L2D$-G    & 0.1144          & 0.1562          & \multicolumn{1}{l|}{\textbf{0.1996}} & \textbf{0.0710} & \textbf{0.0792} & \textbf{0.0862} & 0.0646          & 0.1167          & \multicolumn{1}{l|}{0.1794}          & 0.0295          & 0.0397          & 0.0499          \\
&$L2D$-L    & \textbf{0.1158} & \textbf{0.1569} & \multicolumn{1}{l|}{0.1992}          & 0.0667          & 0.0745          & 0.0813          & \textbf{0.0879} & \textbf{0.1465} & \multicolumn{1}{l|}{\textbf{0.2072}} & \textbf{0.0399} & \textbf{0.0511} & \textbf{0.0596} \\ \hline
\end{tabular}
}
\vspace{-5pt}
\label{table:main}
\end{table*}

\subsection{Item Decoding}

After generating the candidate item representations, $L2D$ could efficiently decode items in the latent space during inference by measuring the similarity between the test sample's hidden state and the representations of the candidate items.
Specifically, for a given test sample with hidden state \(h_t\) and a candidate item \(v\), we denote the candidate item's representation as \(h_v\), which is defined as:  
\begin{equation}
    h_v =
\begin{cases}
\bar{h}_v  \quad \textit{in Eq.~\eqref{eq:global-rep}} & \text{if global aggregation} , \\
\bar{h}_v^t \quad \textit{in Eq.~\eqref{eq:local-rep}} & \text{if local aggregation}.
\end{cases}
\end{equation}
Then, we compute the similarity score between \(h_t\) and \(h_v\) using the L2 distance as:  $S(h_t, h_v) = \frac{1}{\|h_t - h_v\|_2}$.
Once the similarity scores for all candidate items are computed, the top-$K$ items with the highest similarity scores to the test sample are selected to form the final recommendation list. \textbf{We refer to $L2D$ with global aggregation as} $L2D$-G, \textbf{and $L2D$ with local aggregation as} $L2D$-L. 


\section{Experiments}

In this section, we conduct experiments on two widely-used real-world datasets to demonstrate the effectiveness of our $L2D$ framework in balancing performance and inference overhead. 
We will showcase it by following research questions:
\wroll{\textbf{RQ1}: How does the performance and inference overhead of our $L2D$ compare to LLM-based baselines?
\textbf{RQ2}: In which scenarios are global and local aggregation most suitable, respectively? 
\textbf{RQ3}: How does the hyperparameter $M$ affect $L2D$-L?
\textbf{RQ4}: How does the performance of $L2D$ compare to the ID-based classifier?
\textbf{RQ5}: What is the spatial efficiency of $L2D$?
}

\subsection{Experimental Settings}
\noindent \textbf{Datasets.} We evaluated our approach using two representative Amazon Product Review datasets~\footnote{\url{https://jmcauley.ucsd.edu/data/amazon/}}: Amazon CDs (CDs) and Amazon Games (Games). These datasets consist of user review data collected from Amazon between 1996 and 2018. 
Each review was treated as a user-item interaction.
Following~\cite{d3}, we truncated datasets by timestamp to ensure manageable scale, filtered out users/items with fewer than five interactions, and limited user interaction sequences to a maximum length of 10. All interactions was chronologically ordered and splited into training/validation/test sets (8:1:1 ratio). Dataset statistics are detailed in Appendix~\ref{app:Dataset statistics}.

\noindent \textbf{Compared Methods.} 
In this work, to demonstrate the superiority of our proposed method from the perspective of balancing performance and inference overhead in LLM-based recommendation, we primarily selected some of the most commonly used LLM-based models in the current literature. \wroll{\textbf{For LLM-based embedding}, we included \textbf{AlphaRec}~\cite{alpharec} as a baseline. \textbf{For LLM-based generative recommendation}, we included the following methods:
\textbf{BIGRec}~\cite{bigrec}, \textbf{GPT4Rec}~\cite{GPT4Rec}, $\mathbf{D^3}$~\cite{d3}. Additionally, we included \textbf{non-LLM baselines} (\textbf{SASRec}~\cite{sasrec} and \textbf{GRU4Rec}~\cite{gru4rec}) for comprehensive comparison. For a comprehensive description of these baselines, please refer to the Appendix~\ref{app:Compared Methods.}.}

For all generative-based methods, we used beam search to generate multiple items and then map them to real items in the dataset. The implementation details of beam search for recommendation list generation can be found in Appendix~\ref{app:beam}.

\noindent \textbf{Evaluation metrics.}
To evaluate the top-K recommendation performance, we employed two widely adopted metrics: Recall@K and NDCG@K~\cite{d3,zheng2024adapting}. All evaluations follow a full-ranking evaluation protocol~\cite{bigrec}, with K $\in$ \{20, 50, 100\}.
In the following, if space is limited, we will abbreviate Recall@K and NDCG@K as R@K and N@K, respectively.

Other detailed settings are in the appendix~\ref{app:Implementation details}

\subsection{Main Results (RQ1)}

To verify the effectiveness of our $L2D$, we present the performance and inference cost of our method compared to the baseline in Figure~\ref{fig:balance}. Furthermore, we illustrate the performance of our method at different K values in Table~\ref{table:main}.
From the figure and the table, we can find:

\begin{itemize}[leftmargin=*]
\item When evaluating the trade-off between performance and inference cost for all methods, we observe from Figure~\ref{fig:balance} that points closer to the top-left corner indicate better performance at lower costs. Our proposed $L2D$ method is the closest to the top-left corner on both datasets, indicating that $L2D$ achieves excellent performance while maintaining low inference cost, showcasing the effectiveness of \wcb{direct decoding of items in latent space}. Even when compared to the previously most efficient LLM-based method, AlphaRec, which uses LLM as embeddings, $L2D$ reduces the cost by at least a factor of \wcb{five} and gets a better performance, further demonstrating the remarkable potential of $L2D$ in deployment.

\item When comparing the performance of baseline methods under different beam sizes, we observe that the performance of generative-based methods improves approximately linearly as the beam size and inference cost increase. Among these, $D^3$ shows greater scalability (with a larger growth rate). It would not be surprising if these methods could surpass $L2D$ in performance by investing more in inference (e.g., increasing the beam size to 50), but this could lead to nearly a hundredfold increase in cost, which is not feasible in most real-world scenarios. Furthermore, our experiments utilize Llama 3.2-1B as the backbone, which is a relatively small-scale language model. The deployment costs would be even higher with larger language models.

\item Furthermore, Table \ref{table:main} shows the comprehensive performance evaluation.
To rigorously assess the real-world performance of LLM-based models during deployment, inference time costs must be carefully accounted for. Consequently, we adopt the results with the smallest beam size (beam = 1) as the fair comparison baselines for our $L2D$ method, despite the fact that its inference time cost remains substantial (cf. Figure~\ref{fig:balance}). Additionally, we included traditional (i.e. non-LLM) baselines for comprehensive comparison. $L2D$ outperforms all baselines across all metrics. \ww{We attribute this improvement to the method's ability to effectively preserve the powerful generative training characteristics of the LLM's recommendation tuning, thus leveraging the LLM's capabilities acquired during the SFT phase.}


\end{itemize}


\begin{table*}[]
\centering
\caption{$L2D$ vs. ID-based classfier: The overall performance (a) and performance on sparse scenarios (b).}
\renewcommand{\arraystretch}{1.2}
\resizebox{1\linewidth}{!}{
\subfloat[]{
\begin{tabular}{lccc|ccc}
\toprule
\textbf{CDs}        & R@20 & R@50 & R@100 & N@20 & N@50 & N@100 \\ \hline

Classfier    & 0.1087 & 0.1490 & 0.1886  & 0.0634 & 0.0714 & 0.0778  \\
$L2D$  & \textbf{0.1158} & \textbf{0.1569} & \textbf{0.1996}     & \textbf{0.0710} & \textbf{0.0792} & \textbf{0.0862}      \\

\midrule
\textbf{Games}       & R@20 & R@50 & R@100 & N@20 & N@50 & N@100 \\\hline

Classfier    & \textbf{0.0896} & \textbf{0.1557} & \textbf{0.2205}  & 0.0374 & 0.0505 & \textbf{0.0610}  \\
$L2D$  & 0.0879 & 0.1465 & 0.2072     & \textbf{0.0399} & \textbf{0.0511} & 0.0596      \\

\bottomrule
\end{tabular}
}

\hspace{30pt} 
\subfloat[]{
\begin{tabular}{lccc|ccc}
\toprule
\textbf{CDs (Sparse)}        & R@20 & R@50 & R@100 & N@20 & N@50 & N@100 \\ \hline

Classfier    & 0.0491 & 0.0671 & 0.0835  & 0.0271 & 0.0307 & 0.0333   \\
$L2D$  & \textbf{0.0682} & \textbf{0.0889} & \textbf{0.1125}     & \textbf{0.0432} & \textbf{0.0473} & \textbf{0.0511}     \\

\midrule
\textbf{Games (Sparse)}       & R@20 & R@50 & R@100 & N@20 & N@50 & N@100 \\\hline

Classfier    & 0.0242 & 0.0446 & 0.0706  & 0.0095 & 0.0135 & 0.0177  \\
$L2D$  & \textbf{0.0508} & \textbf{0.0874} & \textbf{0.1305}    & \textbf{0.0230} & \textbf{0.0303} & \textbf{0.0372}      \\

\bottomrule
\end{tabular}
}

}

\label{tab:classfier}
\end{table*}


\begin{figure}[t]
  \centering            
  \subfloat[CDs Dense]{
        \label{fig:warmcd}\includegraphics[width=0.22\textwidth]{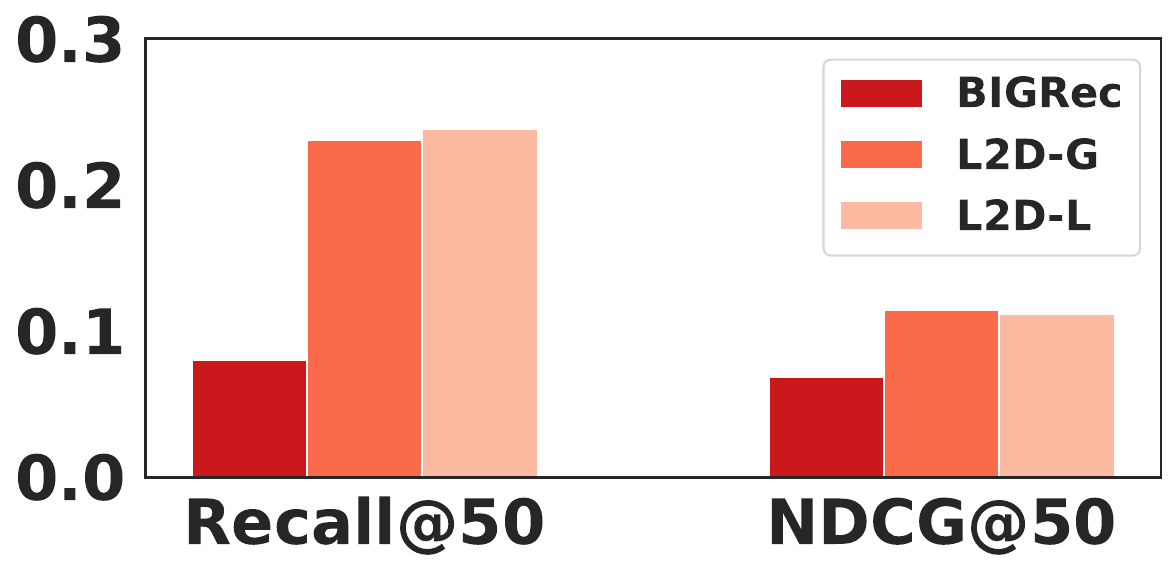}}
  \subfloat[Games Dense]{
        \label{fig:warmgame}\includegraphics[width=0.22\textwidth]{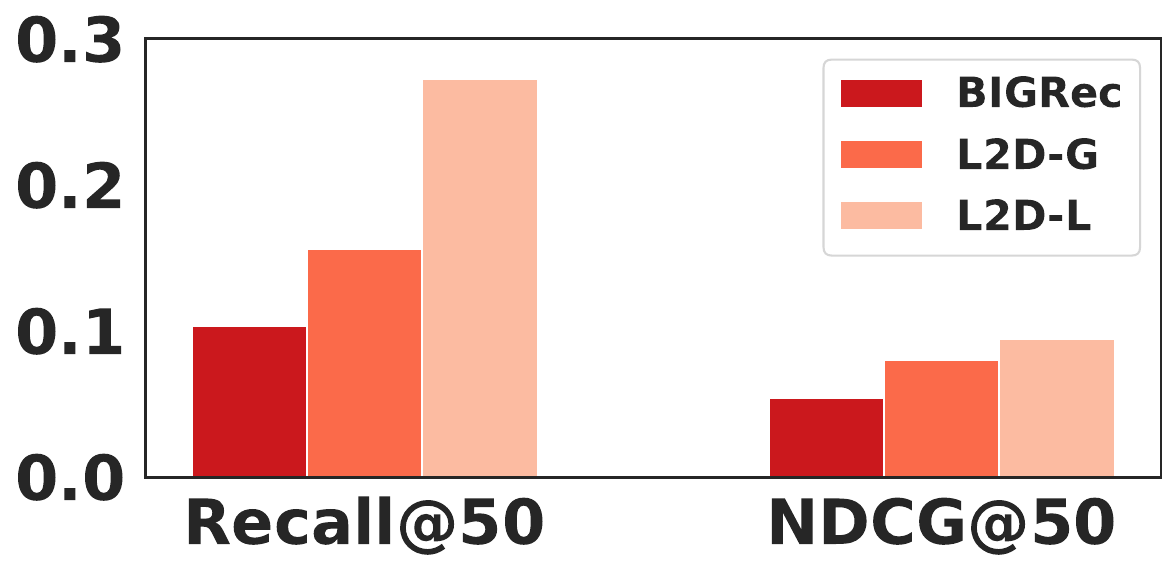}} \\
  \subfloat[CDs Sparse]{
        \label{fig:coldcd}\includegraphics[width=0.22\textwidth]{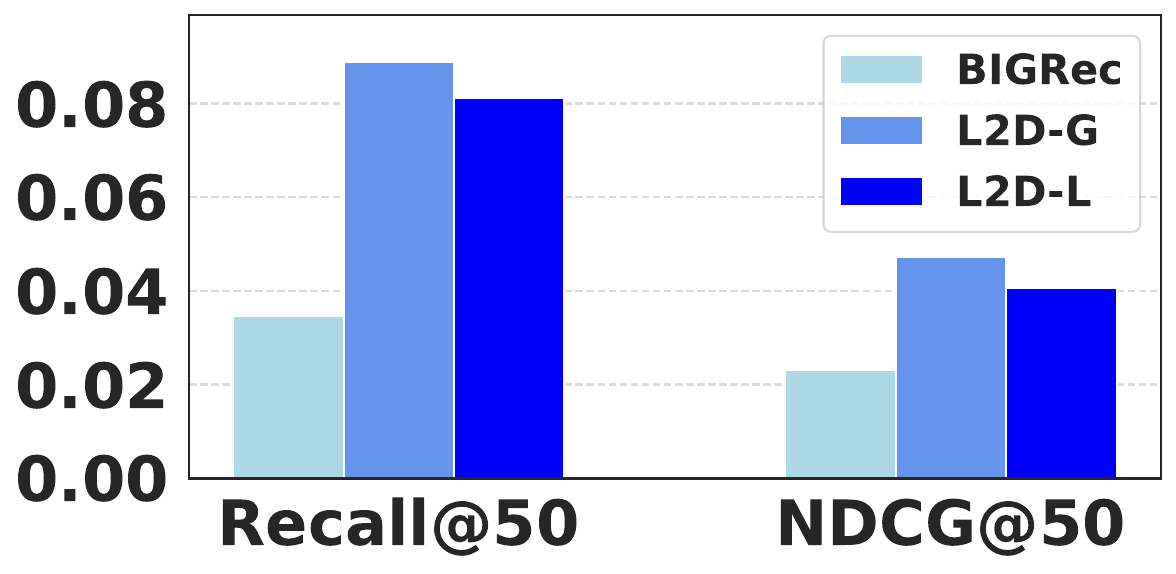}}
  \subfloat[Games Sparse]{
        \label{fig:coldgame}\includegraphics[width=0.22\textwidth]{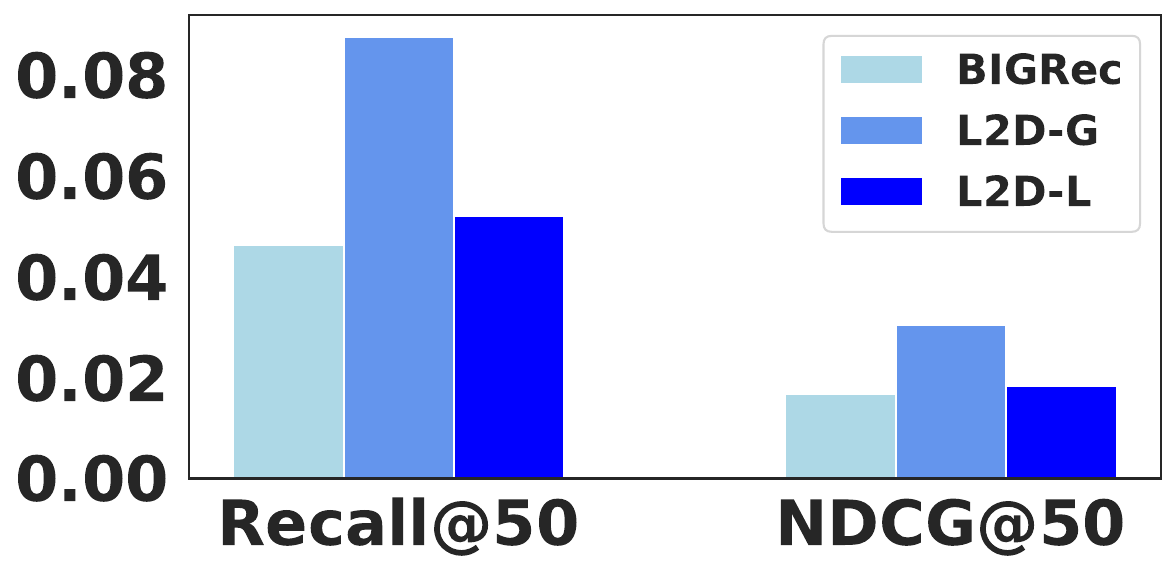}}
  
  \caption{The performance of BIGRec, $L2D$-G, and $L2D$-L on sparse and dense scenarios.}
  \vspace{-20pt}
  \label{fig:cold-warm}
\end{figure}

\subsection{Analysis}


\wroll{In this section, we present a comprehensive analysis of $L2D$. We first discuss its application scenarios in both sparse and dense recommendation settings. Next, we conduct ablation studies to evaluate: 1)The impact of the key hyperparameter M in $L2D$-L, 2)The performance comparison between $L2D$ and ID-based classifiers. Finally, we analyze the space efficiency of the $L2D$ framework.} 

\subsubsection{Sparse and Dense Scenario (RQ2)}
\label{Sparse and Dense}
The discrepancy between $L2D$-G and $L2D$-L results in distinct application scenarios for these two approaches. 
To analyze this, we divided the test set into sparse and dense categories based on item frequency in the training set. Figure~\ref{fig:cold-warm} shows the overall performance of the two strategies in these scenarios. 
We observed the following: (1) \textbf{Dense scenarios:} $L2D$-L achieves the best performance due to the availability of numerous \wcb{hidden states} for each item, allowing it to create a more \wcb{personalized candidate item representation} and eliminate irrelevant information. (2) \textbf{Sparse scenarios:} the interactions are limited, which means that even the top similar \wcb{hidden states in Memory module} may not accurately represent user preferences, potentially leading to biased results and performance drops. In contrast, $L2D$-G, which aggregates preferences globally, offers a more balanced outcome.

\begin{figure}[tbp]
  \centering            
  \subfloat[Recall on CDs]{
        \label{fig:N_cd_r}\includegraphics[width=0.23\textwidth]{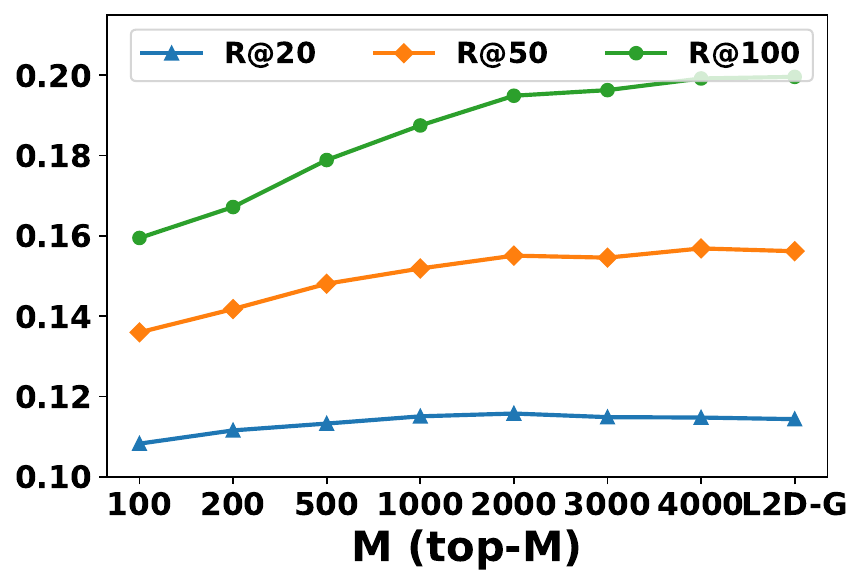}}
  \subfloat[Recall on Games]{
        \label{fig:N_game_r}\includegraphics[width=0.23\textwidth]{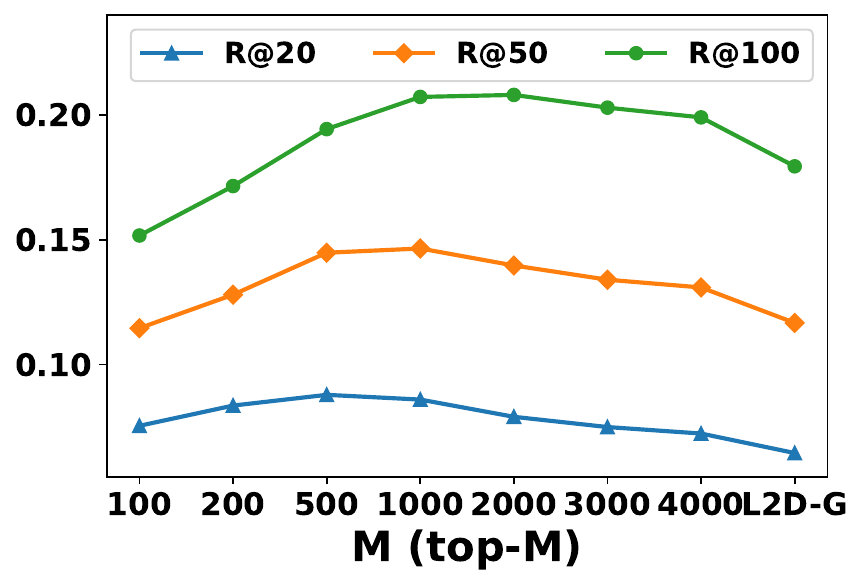}}
        \vspace{-5pt}
  \caption{The impact of $M$ on the Recall metric for $L2D$-L, where $M$ denotes the hyperparameter that determines the number of hidden states in local aggregation. Note that $L2D$-L becomes equivalent to $L2D$-G when $M$ reaches its maximum value.}
  \label{fig:Mimpact}
\end{figure}

\begin{table}[]
\centering
\caption{The performance of $L2D$ when storing only 30\% of the training samples. The best results are in bold.}
\renewcommand{\arraystretch}{1.2}
\resizebox{1.\linewidth}{!}{

\begin{tabular}{lccc|ccc}
\toprule
\textbf{CDs}       & R@20 & R@50 & R@100 & N@20 & N@50 & N@100 \\ \hline
SASRec   & \textbf{0.1015} & 0.1271 & 0.1522  & 0.0602 & 0.0653 & 0.0693   \\
GRU4Rec  & 0.0707 & 0.1027 & 0.1347 & 0.0376 & 0.0439 & 0.0491   \\ 

AlphaRec    & 0.0651 & 0.0976 & 0.1353 & 0.0300 & 0.0364 & 0.0425 \\
GPT4Rec (beam=1)    & 0.0513 & 0.0562 & 0.0652 & 0.0433 & 0.0443 & 0.0458 \\
BIGRec (beam=1)     & 0.0506 & 0.0565 & 0.0621 & 0.0435 & 0.0446 & 0.0456 \\
$D^3$ (beam=1)      & 0.0507 & 0.0560 & 0.0623 & 0.0436 & 0.0447 & 0.0457 \\\hline
$L2D$-G (30\%) & 0.1012 & \textbf{0.1391} & \textbf{0.1807} & \textbf{0.0621} & \textbf{0.0696} & \textbf{0.0763} \\
$L2D$-L (30\%) & 0.0991 & 0.1344 & 0.1678 & 0.0579 & 0.0649 & 0.0703 \\

\midrule
\textbf{Games}      & R@20 & R@50 & R@100 & N@20 & N@50 & N@100 \\\hline
SASRec    & 0.0684 & 0.1117 & 0.1564 & 0.0332 &0.0417 & 0.0490  \\
GRU4Rec   & 0.0664 & 0.1099 & 0.1601 & 0.0302 & 0.0387 & 0.0468  \\ 
AlphaRec    & 0.0619  & 0.1005  & 0.1392  & 0.0295  & 0.0371  & 0.0434  \\
GPT4Rec (beam=1)    & 0.0508  & 0.0782  & 0.1064  & 0.0293  & 0.0347  & 0.0392  \\
BIGRec (beam=1)     & 0.0476  & 0.0702  & 0.1007  & 0.0284  & 0.0328  & 0.0378  \\
$D^3$ (beam=1)      & 0.0478  & 0.0711  & 0.1004  & 0.0284  & 0.033   & 0.0376  \\ \hline
$L2D$-G (30\%) & \textbf{0.0759} & \textbf{0.1314} & \textbf{0.1973} & \textbf{0.0341} & \textbf{0.0450} & \textbf{0.0556} \\
$L2D$-L (30\%) & 0.0682 & 0.1214 & 0.1795 & 0.0305 & 0.0407 & 0.0500 \\
\bottomrule
\end{tabular}
}
\label{tab:space}
\end{table}

\subsubsection{Hyper-parameter $M$ on $L2D$-L (RQ3)}

\roll{We illustrate the impact of $M$ in Figure~\ref{fig:Mimpact}, where only the results for Recall are reported. The results for NDCG can be found in Appendix~\ref{app:Impact_ndcg}}.

Specifically, on the CDs dataset, the Global Aggregation method in $L2D$-G outperforms the Local Aggregation method in $L2D$-L. In contrast, on the Games dataset, we observe that performance peaks as $M$ increases, but further increasing $M$ leads to a decline in performance. We attribute this phenomenon to the varying demands for focusing on the test sample's feature aspects in different recommendation scenarios. The Games dataset may require a stronger emphasis on detailed feature aspects compared to the CDs dataset.

\wroll{
\subsubsection{L2D vs. ID-based classfier (RQ4)}
\label{classfier}
Training a recommendation head to decode the next item by directly predicting its ID provides a straightforward way to avoid language-space decoding, improving decoding efficiency. However, this \ww{training} objective deviates from the LLM’s original goal of next-token prediction, which may hinder effective utilization of its pretrained knowledge. Moreover, this approach incurs additional training overhead and may perform poorly on sparse items, as it requires learning in the LLM’s high-dimensional representation space.  In contrast, our method can bypass language-space decoding while still preserving the \ww{powerful} generative training characteristics of the \ww{LLM's} recommendation tuning, performing well on sparse items recommendation scenarios.

To verify this, we trained a classifier head using the LLM's hidden states with careful hyperparameter tuning. The results, summarized in Table~\ref{tab:classfier}, report both (a) overall performance and (b) performance on sparse recommendation scenario. For overall performance (a), on the CDs dataset, our method consistently outperforms the baseline, achieving an average relative improvement of 11.2\%. On the Games dataset, our method performs better for smaller K in NDCG and remains comparable for larger K in both metrics. Regarding performance on sparse recommendation scenario (b), our method consistently achieves significantly better results.
}
%

\wroll{
\subsubsection{Spatial complexity (RQ5)}
\label{space}

In this subsection, we analyze the spatial complexity of the proposed $L2D$ framework. Although the pre-stored hidden states of training samples in $L2D$ introduce additional space requirements, these costs remain manageable since this scale of storage is feasible even for personal devices. For instance, if each sample corresponds to a 1024-dimensional hidden state (float16), storing hidden states for $10^9$ training samples requires approximately 1024×2 bytes × $10^9 \approx $ 2 TB. 

Storage costs can be further optimized by selectively retaining only a subset of training samples, such as through reservoir sampling technique~\cite{DBLP:journals/corr/abs-2403-03993}, where new data is added over time while older data is removed. Our experiments show that even when storing only 30\% of the orignal training data, our method still outperforms baselines (where only the Recall@20 metric on the CDs dataset is competitive with SASRec). The results are shown in Table~\ref{tab:space}.


}

\section{Related Work}




\noindent $\bullet$ \textbf{LLM-based recommendation.} Large language models have been extensively employed in diverse personalization tasks~\cite{dpl,nextquill,zhao2025exploring,shen2024pmg,xu2025personalized,qiu2025latent}, where recommendation constitutes an important application. Here, we discuss three paradigms of LLM-based recommenders~\cite{wu2024survey}. (1) \textbf{LLM-Embedding-Based Recommenders} use embeddings from LLMs in traditional systems to capture user preferences~\cite{Re_11,Re_12}. While effective in language tasks, these embeddings require fine-tuning for optimal performance. (2) \textbf{LLM-Based Discriminative Recommenders} directly predict user-item interactions by optimizing the recommendation task with the LLM's loss function~\cite{Re_21,Re_22,Re_23}. Although it dispenses with intermediate embeddings, it requires evaluating each item individually, reducing efficiency compared to traditional models~\cite{sun2022enhancing}. (3) \textbf{LLM-Based Generative Recommenders} generate natural language recommendations without predefined items, offering innovative potential~\cite{bigrec,d3,zheng2024adapting,rebuttal1,rebuttal2,zhang2025reinforced}. However, autoregressive decoding introduces significant inference overhead. 
Inspired by these paradigms, we propose a novel LLM-based recommender that balances performance and overhead, addressing existing challenges to enhance quality and efficiency. 


\label{contriution add}
\roll{
Notably, some existing (large) language model (LM)-based approaches~\cite{alpharec}, such as RecFormer~\cite{DBLP:conf/kdd/LiWLFSSM23}, can be viewed as representing candidate items in latent spaces and then matching them with the user input sequence encoded by the LM. 
However, they indeed modify the output layer of the LMs, with the effectiveness of their matching process tied to the training process. As a result, they fail to achieve plug-and-play integration into existing advanced LLM-based recommenders. In contrast, our method is decoupled from the training process, making it plug-and-play. Additionally, these methods' training objectives deviate from the large language model’s original goal of next-token prediction, which may hinder effective utilization of its pretrained knowledge. 
A detailed discussion is provided in Appendix~\ref{app:positioning}.
}

\noindent $\bullet$ \textbf{Inference Acceleration for LLM-based Recommendation.}
With the widespread application of LLMs~\cite{liu2025inference,nextquill,zhao2024pacar}, an increasing number of studies have focused on accelerating LLM inference. In particular, in the field of LLM-based recommender systems, models need to recommend products to a large number of users within a short time frame, which highlights the necessity of considering methods to accelerate LLM inference in this domain. Speculative Decoding (SD) \cite{Re_sd}, a significant acceleration technique in the NLP field, has been applied to recommender systems, such as DARE~\cite{Re_41} and AtSpeed~\cite{Re_42}. 
However, these methods still rely on acceleration decoding within the language space. In contrast, our method takes a step further by exploring how to implement efficient decoding for recommendation in the latent space of LLMs, while maintaining a simple and easy-to-implement overall framework that avoids complex designs.

\section{Conclusion}



In this study, we emphasized that fine-tuning LLMs for recommendations in a generative manner is highly promising but encounters significant inference overhead due to the original autoregressive decoding strategy. To address this challenge, we proposed the $L2D$, which bypasses time-consuming autoregressive decoding in the language space and directly decodes items in LLM's latent space. 
The $L2D$ preserves the generative tuning paradigm to keep performance while enabling efficient decoding.
Our results highlighted the potential of latent space decoding as a fundamental advancement in LLM-based recommender systems, and extensive results demonstrated the superiority of $L2D$.

\section*{Limitations}


This paper has the following limitations: 1) Although the $L2D$ framework we introduced significantly reduces inference latency, the memory, which is pre-constructed, still incurs additional time overhead during its pre-construction process. This motivates us to explore more efficient memory construction methods in future work. 2) 
\wroll{Our approach, while capable of processing items with at least one interaction without requiring additional training (unlike traditional methods such as SASRec/GRU4Rec and some LLM-based methods such as D3/ID-based classfier that need retraining), still shares the same fundamental limitation as conventional methods: it cannot handle fully cold-start items with zero interaction history. In the future, we plan to address this issue by using the interpolation technique or incorporating auxiliary models. }
3) We have not considered the problem of memory updating. As user interaction data gradually accumulates over time, how to effectively use this new data to update the memory in $L2D$ to achieve higher decoding performance presents a promising direction. We intend to explore this issue in future research. Besides, we will also examine the effectiveness of our method when combined with other personalized preference alignment approaches in training (e.g., causality-aware approach~\cite{nextquill, Re_33}).
\section*{Ethical Considerations}

In this paper, we present $L2D$, designed to balance the performance and inference overhead for generative LLMRec. Our method decode item in latent space of LLM which doesn’t raise ethical concerns. Moreover, the data we use are publicly available and don’t include sensitive details. However, recommendations involve user
behavioral data, which might raise privacy concerns, which can be addressed through introducing the mechanism of user consent. Additionally, using
LLMs may have potential negative societal biases. We argue for a thorough risk assessment and alert users to the potential risks associated with model deployment.

For the large language model use,
we utilize ChatGPT to help polish the writing at the sentence level.

\section*{Acknowledgments}
This work is supported by the National Natural Science Foundation of China (62272437) and the advanced computing resources provided by the Supercomputing Center of the USTC.

\bibliography{latex/main/9_bib}

\begin{thebibliography}{34}
\providecommand{\natexlab}[1]{#1}

\bibitem[{Bao et~al.(2023{\natexlab{a}})Bao, Zhang, Wang, Zhang, Yang, Luo, Chen, Feng, and Tian}]{bigrec}
Keqin Bao, Jizhi Zhang, Wenjie Wang, Yang Zhang, Zhengyi Yang, Yancheng Luo, Chong Chen, Fuli Feng, and Qi~Tian. 2023{\natexlab{a}}.
\newblock A bi-step grounding paradigm for large language models in recommendation systems.
\newblock \emph{arXiv preprint arXiv:2308.08434}.

\bibitem[{Bao et~al.(2024)Bao, Zhang, Zhang, Huo, Chen, and Feng}]{d3}
Keqin Bao, Jizhi Zhang, Yang Zhang, Xinyue Huo, Chong Chen, and Fuli Feng. 2024.
\newblock \href {https://aclanthology.org/2024.emnlp-main.589} {Decoding matters: Addressing amplification bias and homogeneity issue in recommendations for large language models}.
\newblock In \emph{Proceedings of the 2024 Conference on Empirical Methods in Natural Language Processing}, pages 10540--10552.

\bibitem[{Bao et~al.(2023{\natexlab{b}})Bao, Zhang, Zhang, Wang, Feng, and He}]{tallrec}
Keqin Bao, Jizhi Zhang, Yang Zhang, Wenjie Wang, Fuli Feng, and Xiangnan He. 2023{\natexlab{b}}.
\newblock \href {https://doi.org/10.1145/3604915.3608857} {Tallrec: An effective and efficient tuning framework to align large language model with recommendation}.
\newblock In \emph{Proceedings of the 17th ACM Conference on Recommender Systems}, pages 1007--1014.

\bibitem[{Dubey et~al.(2024)Dubey, Jauhri, Pandey, Kadian, Al-Dahle, Letman, Mathur, Schelten, Yang, Fan et~al.}]{LLAMA3}
Abhimanyu Dubey, Abhinav Jauhri, Abhinav Pandey, Abhishek Kadian, Ahmad Al-Dahle, Aiesha Letman, Akhil Mathur, Alan Schelten, Amy Yang, Angela Fan, et~al. 2024.
\newblock The llama 3 herd of models.
\newblock \emph{arXiv preprint arXiv:2407.21783}.

\bibitem[{Hidasi et~al.(2016)Hidasi, Karatzoglou, Baltrunas, and Tikk}]{gru4rec}
Bal{\'{a}}zs Hidasi, Alexandros Karatzoglou, Linas Baltrunas, and Domonkos Tikk. 2016.
\newblock Session-based recommendations with recurrent neural networks.
\newblock In \emph{{ICLR} (Poster)}.

\bibitem[{Hua et~al.(2023)Hua, Xu, Ge, and Zhang}]{rebuttal1}
Wenyue Hua, Shuyuan Xu, Yingqiang Ge, and Yongfeng Zhang. 2023.
\newblock How to index item ids for recommendation foundation models.
\newblock In \emph{{SIGIR-AP}}, pages 195--204. {ACM}.

\bibitem[{Kang and McAuley(2018)}]{sasrec}
Wang{-}Cheng Kang and Julian~J. McAuley. 2018.
\newblock Self-attentive sequential recommendation.
\newblock In \emph{{ICDM}}, pages 197--206. {IEEE} Computer Society.

\bibitem[{Leviathan et~al.(2023)Leviathan, Kalman, and Matias}]{Re_sd}
Yaniv Leviathan, Matan Kalman, and Yossi Matias. 2023.
\newblock Fast inference from transformers via speculative decoding.
\newblock In \emph{{ICML}}, volume 202 of \emph{Proceedings of Machine Learning Research}, pages 19274--19286. {PMLR}.

\bibitem[{Li et~al.(2023{\natexlab{a}})Li, Wang, Li, Fu, Shen, Shang, and McAuley}]{DBLP:conf/kdd/LiWLFSSM23}
Jiacheng Li, Ming Wang, Jin Li, Jinmiao Fu, Xin Shen, Jingbo Shang, and Julian~J. McAuley. 2023{\natexlab{a}}.
\newblock Text is all you need: Learning language representations for sequential recommendation.
\newblock In \emph{{KDD}}, pages 1258--1267. {ACM}.

\bibitem[{Li et~al.(2023{\natexlab{b}})Li, Chen, Zhao, Zhang, and Xing}]{Re_22}
Xinhang Li, Chong Chen, Xiangyu Zhao, Yong Zhang, and Chunxiao Xing. 2023{\natexlab{b}}.
\newblock E4srec: An elegant effective efficient extensible solution of large language models for sequential recommendation.
\newblock \emph{CoRR}, abs/2312.02443.

\bibitem[{Lin et~al.(2024{\natexlab{a}})Lin, Yang, Wang, Li, Du, Feng, Ng, and Chua}]{lin2024efficient}
Xinyu Lin, Chaoqun Yang, Wenjie Wang, Yongqi Li, Cunxiao Du, Fuli Feng, See-Kiong Ng, and Tat-Seng Chua. 2024{\natexlab{a}}.
\newblock Efficient inference for large language model-based generative recommendation.
\newblock \emph{arXiv preprint arXiv:2410.05165}.

\bibitem[{Lin et~al.(2024{\natexlab{b}})Lin, Yang, Wang, Li, Du, Feng, Ng, and Chua}]{Re_42}
Xinyu Lin, Chaoqun Yang, Wenjie Wang, Yongqi Li, Cunxiao Du, Fuli Feng, See{-}Kiong Ng, and Tat{-}Seng Chua. 2024{\natexlab{b}}.
\newblock Efficient inference for large language model-based generative recommendation.
\newblock \emph{CoRR}, abs/2410.05165.

\bibitem[{Liu et~al.(2025)Liu, Du, Zhao, Zhang, Zhao, Wang, Dong, and Xu}]{liu2025inference}
Weihao Liu, Zhaocheng Du, Haiyuan Zhao, Wenbo Zhang, Xiaoyan Zhao, Gang Wang, Zhenhua Dong, and Jun Xu. 2025.
\newblock Inference computation scaling for feature augmentation in recommendation systems.
\newblock \emph{arXiv preprint arXiv:2502.16040}.

\bibitem[{Qiu et~al.(2025{\natexlab{a}})Qiu, Shi, Zhao, Zhu, Zhang, and Feng}]{qiu2025latent}
Yilun Qiu, Tianhao Shi, Xiaoyan Zhao, Fengbin Zhu, Yang Zhang, and Fuli Feng. 2025{\natexlab{a}}.
\newblock Latent inter-user difference modeling for llm personalization.
\newblock \emph{arXiv preprint arXiv:2507.20849}.

\bibitem[{Qiu et~al.(2025{\natexlab{b}})Qiu, Zhao, Zhang, Bai, Wang, Cheng, Feng, and Chua}]{dpl}
Yilun Qiu, Xiaoyan Zhao, Yang Zhang, Yimeng Bai, Wenjie Wang, Hong Cheng, Fuli Feng, and Tat-Seng Chua. 2025{\natexlab{b}}.
\newblock \href {https://doi.org/10.18653/v1/2025.findings-acl.1095} {Measuring what makes you unique: Difference-aware user modeling for enhancing {LLM} personalization}.
\newblock In \emph{Findings of the Association for Computational Linguistics: ACL 2025}, pages 21258--21277, Vienna, Austria. Association for Computational Linguistics.

\bibitem[{Shen et~al.(2024)Shen, Zhang, Zhao, Zhu, and Xiao}]{shen2024pmg}
Xiaoteng Shen, Rui Zhang, Xiaoyan Zhao, Jieming Zhu, and Xi~Xiao. 2024.
\newblock Pmg: Personalized multimodal generation with large language models.
\newblock In \emph{Proceedings of the ACM Web Conference 2024}, pages 3833--3843.

\bibitem[{Sheng et~al.(2025)Sheng, Zhang, Zhang, Chen, Wang, and Chua}]{alpharec}
Leheng Sheng, An~Zhang, Yi~Zhang, Yuxin Chen, Xiang Wang, and Tat{-}Seng Chua. 2025.
\newblock Language representations can be what recommenders need: Findings and potentials.
\newblock In \emph{{ICLR}}. OpenReview.net.

\bibitem[{Sun et~al.(2022)Sun, Yuan, Yang, Karatzoglou, Shen, and Zhao}]{sun2022enhancing}
Yang Sun, Fajie Yuan, Min Yang, Alexandros Karatzoglou, Li~Shen, and Xiaoyan Zhao. 2022.
\newblock Enhancing top-n item recommendations by peer collaboration.
\newblock In \emph{Proceedings of the 45th International ACM SIGIR Conference on Research and Development in Information Retrieval}, pages 1895--1900.

\bibitem[{Valkanas et~al.(2024)Valkanas, Wang, Zhang, and Coates}]{DBLP:journals/corr/abs-2403-03993}
Antonios Valkanas, Yuening Wang, Yingxue Zhang, and Mark Coates. 2024.
\newblock Personalized negative reservoir for incremental learning in recommender systems.
\newblock \emph{CoRR}, abs/2403.03993.

\bibitem[{Wu et~al.(2024)Wu, Zheng, Qiu, Wang, Gu, Shen, Qin, Zhu, Zhu, Liu et~al.}]{wu2024survey}
Likang Wu, Zhi Zheng, Zhaopeng Qiu, Hao Wang, Hongchao Gu, Tingjia Shen, Chuan Qin, Chen Zhu, Hengshu Zhu, Qi~Liu, et~al. 2024.
\newblock A survey on large language models for recommendation.
\newblock \emph{World Wide Web}, 27(5):60.

\bibitem[{Xi et~al.(2024{\natexlab{a}})Xi, Liu, Lin, Cai, Zhu, Zhu, Chen, Tang, Zhang, and Yu}]{Re_12}
Yunjia Xi, Weiwen Liu, Jianghao Lin, Xiaoling Cai, Hong Zhu, Jieming Zhu, Bo~Chen, Ruiming Tang, Weinan Zhang, and Yong Yu. 2024{\natexlab{a}}.
\newblock Towards open-world recommendation with knowledge augmentation from large language models.
\newblock In \emph{RecSys}, pages 12--22. {ACM}.

\bibitem[{Xi et~al.(2024{\natexlab{b}})Xi, Wang, Chen, Lin, Zhu, Liu, Tang, Zhang, and Yu}]{Re_41}
Yunjia Xi, Hangyu Wang, Bo~Chen, Jianghao Lin, Menghui Zhu, Weiwen Liu, Ruiming Tang, Weinan Zhang, and Yong Yu. 2024{\natexlab{b}}.
\newblock A decoding acceleration framework for industrial deployable llm-based recommender systems.
\newblock \emph{CoRR}, abs/2408.05676.

\bibitem[{Xu et~al.(2025)Xu, Wang, Zhang, Tang, Yan, Feng, and He}]{xu2025personalized}
Yiyan Xu, Wenjie Wang, Yang Zhang, Biao Tang, Peng Yan, Fuli Feng, and Xiangnan He. 2025.
\newblock Personalized image generation with large multimodal models.
\newblock In \emph{Proceedings of the ACM on Web Conference 2025}, pages 264--274.

\bibitem[{Yuan et~al.(2023)Yuan, Yuan, Song, Li, Fu, Yang, Pan, and Ni}]{Re_11}
Zheng Yuan, Fajie Yuan, Yu~Song, Youhua Li, Junchen Fu, Fei Yang, Yunzhu Pan, and Yongxin Ni. 2023.
\newblock Where to go next for recommender systems? {ID-} vs. modality-based recommender models revisited.
\newblock In \emph{{SIGIR}}, pages 2639--2649. {ACM}.

\bibitem[{Zhang et~al.(2024{\natexlab{a}})Zhang, Yan, Zhang, Kang, Li, Huang, Wang, and Kim}]{GPT4Rec}
Peiyan Zhang, Yuchen Yan, Xi~Zhang, Liying Kang, Chaozhuo Li, Feiran Huang, Senzhang Wang, and Sunghun Kim. 2024{\natexlab{a}}.
\newblock Gpt4rec: Graph prompt tuning for streaming recommendation.
\newblock In \emph{{SIGIR}}, pages 1774--1784. {ACM}.

\bibitem[{Zhang et~al.(2024{\natexlab{b}})Zhang, Bao, Yan, Wang, Feng, and He}]{Re_23}
Yang Zhang, Keqin Bao, Ming Yan, Wenjie Wang, Fuli Feng, and Xiangnan He. 2024{\natexlab{b}}.
\newblock Text-like encoding of collaborative information in large language models for recommendation.
\newblock In \emph{{ACL} {(1)}}, pages 9181--9191. Association for Computational Linguistics.

\bibitem[{Zhang et~al.(2025{\natexlab{a}})Zhang, Feng, Zhang, Bao, Wang, and He}]{Re_21}
Yang Zhang, Fuli Feng, Jizhi Zhang, Keqin Bao, Qifan Wang, and Xiangnan He. 2025{\natexlab{a}}.
\newblock Collm: Integrating collaborative embeddings into large language models for recommendation.
\newblock \emph{IEEE Transactions on Knowledge and Data Engineering}.

\bibitem[{Zhang et~al.(2025{\natexlab{b}})Zhang, Xu, Zhao, Wang, Feng, He, and Chua}]{zhang2025reinforced}
Yang Zhang, Wenxin Xu, Xiaoyan Zhao, Wenjie Wang, Fuli Feng, Xiangnan He, and Tat-Seng Chua. 2025{\natexlab{b}}.
\newblock Reinforced latent reasoning for llm-based recommendation.
\newblock \emph{arXiv preprint arXiv:2505.19092}.

\bibitem[{Zhang et~al.(2024{\natexlab{c}})Zhang, You, Bai, Zhang, Bao, Wang, and Chua}]{Re_33}
Yang Zhang, Juntao You, Yimeng Bai, Jizhi Zhang, Keqin Bao, Wenjie Wang, and Tat{-}Seng Chua. 2024{\natexlab{c}}.
\newblock Causality-enhanced behavior sequence modeling in llms for personalized recommendation.
\newblock \emph{CoRR}, abs/2410.22809.

\bibitem[{Zhao et~al.(2025{\natexlab{a}})Zhao, Deng, Wang, Cheng, Zhang, Ng, Chua et~al.}]{zhao2025exploring}
Xiaoyan Zhao, Yang Deng, Wenjie Wang, Hong Cheng, Rui Zhang, See-Kiong Ng, Tat-Seng Chua, et~al. 2025{\natexlab{a}}.
\newblock Exploring the impact of personality traits on conversational recommender systems: A simulation with large language models.
\newblock \emph{arXiv preprint arXiv:2504.12313}.

\bibitem[{Zhao et~al.(2024)Zhao, Wang, Wang, Cheng, Zhang, and Wong}]{zhao2024pacar}
Xiaoyan Zhao, Lingzhi Wang, Zhanghao Wang, Hong Cheng, Rui Zhang, and Kam-Fai Wong. 2024.
\newblock Pacar: Automated fact-checking with planning and customized action reasoning using large language models.
\newblock In \emph{Proceedings of the 2024 Joint International Conference on Computational Linguistics, Language Resources and Evaluation (LREC-COLING 2024)}, pages 12564--12573.

\bibitem[{Zhao et~al.(2025{\natexlab{b}})Zhao, You, Zhang, Wang, Cheng, Feng, Ng, and Chua}]{nextquill}
Xiaoyan Zhao, Juntao You, Yang Zhang, Wenjie Wang, Hong Cheng, Fuli Feng, See-Kiong Ng, and Tat-Seng Chua. 2025{\natexlab{b}}.
\newblock Nextquill: Causal preference modeling for enhancing llm personalization.
\newblock \emph{arXiv preprint arXiv:2506.02368}.

\bibitem[{Zheng et~al.(2024)Zheng, Hou, Lu, Chen, Zhao, Chen, and Wen}]{zheng2024adapting}
Bowen Zheng, Yupeng Hou, Hongyu Lu, Yu~Chen, Wayne~Xin Zhao, Ming Chen, and Ji-Rong Wen. 2024.
\newblock Adapting large language models by integrating collaborative semantics for recommendation.
\newblock In \emph{2024 IEEE 40th International Conference on Data Engineering (ICDE)}, pages 1435--1448. IEEE.

\bibitem[{Zhu et~al.(2024)Zhu, Wu, Guo, Hong, and Li}]{rebuttal2}
Yaochen Zhu, Liang Wu, Qi~Guo, Liangjie Hong, and Jundong Li. 2024.
\newblock Collaborative large language model for recommender systems.
\newblock In \emph{{WWW}}, pages 3162--3172. {ACM}.

\end{thebibliography}
\clearpage
\newpage
\appendix

\section{Appendix}

\subsection{Contribution Positioning}
\label{app:positioning}

Some existing (large) language model (LM)-based approaches~\cite{alpharec}, such as RecFormer~\cite{DBLP:conf/kdd/LiWLFSSM23}, can be seen as representing candidate items in latent spaces and matching them with the user input sequence encoded by the LM. This makes them somewhat similar to our approach. However, there are inherent differences between these methods and ours. First, our method does not alter the generative training process (next-token prediction); it only modifies the decoding process without requiring additional tuning. In contrast, in these existing approaches, the matching process is entangled with the training phase. As a result, they fail to achieve plug-and-play integration into existing advanced LLM-based recommenders. Also, the training objective of these methods deviates from the language model’s original goal of next-token prediction, which may hinder effective utilization of its pretrained knowledge. 

Secondly, even when focusing solely on the matching process, there are differences in how the sequence representations and candidate item representations are constructed, as well as in the learning processes involved. Our approach introduces the following innovations:
\begin{itemize}[leftmargin=*, itemsep=0pt,parsep=1pt]
    \item History Representation: Our representation is derived from the hidden state embedding at the next-token prediction position, which serves as a "generative state" inherently encoding information for generating subsequent tokens. In contrast, the existing methods do not leverage such a generative state of LLMs.

    \item Item Representation: We construct item representations by aggregating the "generative states" of training samples where the item appears as the target. This fundamentally differs from existing works, which require an item-based forward encoding approach.
    
    \item Learning: Our history and item representations exist in the same space and do not require additional tuning. In contrast, existing methods necessitate a separate training process to align these representations for matching.
\end{itemize}

\subsection{Beam-search for Recommendation}
\label{app:beam}
For all generative-based methods, we use beam search to generate multiple items and then match them to real items. Specifically, we first obtain the semantic representation of each generated item and compute their matching scores based on their semantic similarity with all candidate items. This results in a ranking matrix with dimensions beam\_number × candidate\_item\_number, where each row represents the ranking list of a beam-generated item. Finally, we flatten the matrix column by column into a single vector and retain the top K unique items as the recommendation results.

\subsection{Implementation details}
\label{app:Implementation details}
Our LLM-based recommendation models were built on Llama3.2-1B~\cite{LLAMA3} as the backbone architecture. The performance of our method on Llama3.1-8B can be found in the Appendix~\ref{Generalizability}. During the instruction tuning phase, we adopted the AdamW optimizer along with a cosine learning rate scheduler, using a batch size of 64 and exploring learning rates in \{1e-3, 1e-4, 5e-5\}. Other hyperparameters align with the default configurations from the $D^3$ paper~\cite{d3}. All experiments are conducted on NVIDIA A100 GPUs.

\subsection{Generalizability}
\label{Generalizability}
To further show the generalizability of our approach, we evaluate its performance on Llama3.1-8B (Table~\ref{tab:llama8B}). The results demonstrate the robust scalability of our approach and highlight its applicability to larger backbone models.

\begin{table*}[]

\centering
\caption{Performance comparison on the CDs and Games datasets using Llama3.1-8B. The best results are in bold.}
\small
\begin{tabular}{llcccccc}
\toprule
\textbf{Dataset} & \textbf{Model} & \textbf{R@20} & \textbf{R@50} & \textbf{R@100} & \textbf{N@20} & \textbf{N@50} & \textbf{N@100} \\
\midrule
\multirow{4}{*}{CDs} 
 & BIGRec-1B (beam=1) & 0.0506 & 0.0565 & 0.0621 & 0.0435 & 0.0446 & 0.0456 \\
 & BIGRec-8B (beam=1) & 0.0514 & 0.0573 & 0.0620 & 0.0457 & 0.0469 & 0.0477 \\
 & L2D-1B             & 0.1158 & 0.1569 & 0.1996 & 0.0710 & 0.0792 & 0.0862 \\
 & L2D-8B             & \textbf{0.1243} & \textbf{0.1654} & \textbf{0.2113} & \textbf{0.0770} & \textbf{0.0851} & \textbf{0.0925} \\
\midrule
\multirow{4}{*}{Games} 
 & BIGRec-1B (beam=1) & 0.0476 & 0.0702 & 0.1007 & 0.0284 & 0.0328 & 0.0378 \\
 & BIGRec-8B (beam=1) & 0.0524 & 0.0763 & 0.1023 & 0.0337 & 0.0384 & 0.0426 \\
 & L2D-1B             & 0.0879 & 0.1465 & 0.2072 & 0.0399 & 0.0511 & 0.0596 \\
 & L2D-8B             & \textbf{0.0936} & \textbf{0.1575} & \textbf{0.2240} & \textbf{0.0416} & \textbf{0.0537} & \textbf{0.0637} \\
\bottomrule
\end{tabular}
\label{tab:llama8B}
\end{table*}

\subsection{Dataset Statistics}
\label{app:Dataset statistics}

In this subsection, we supplement the statistical information of the datasets used in our experiments. Please refer to Table~\ref{table:data}

\begin{table}[tbp]
    \centering
    \renewcommand{\arraystretch}{1.2}
    \caption{The statistics of datasets.}
    \resizebox{1\linewidth}{!}{
    \begin{tabular}{ccccccc}
    \hline
        Dataset & \#User & \#Item & \#Train & \#Valid & \#Test\\
        \hline
        CDs &  21,347 & 14,239 &  148,685 & 18,586 & 18,587\\
        Games & 34,089 & 11,037 & 201,613 &  25,202 &  25,203\\
        \hline
    \end{tabular}
}
    \label{table:data}
    \vspace{-1em}
\end{table}

\subsection{Impact of $M$ on NDCG for L2D-L}
\label{app:Impact_ndcg}
\begin{figure}[tbp]
  \centering            
  \subfloat[CDs NDCG]{
        \label{fig:N_cd_n}\includegraphics[width=0.24\textwidth]{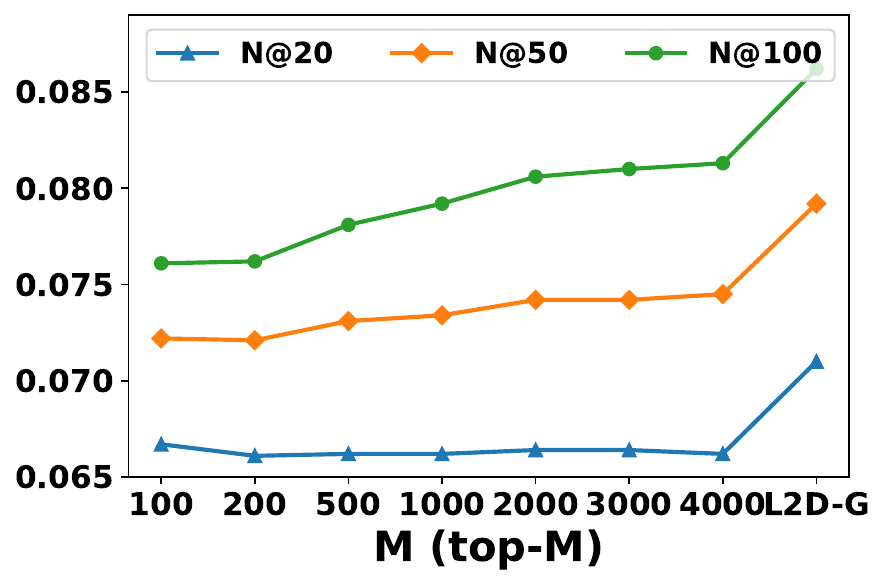}} 
  \subfloat[Games NDCG]{
        \label{fig:N_game_n}\includegraphics[width=0.235\textwidth]{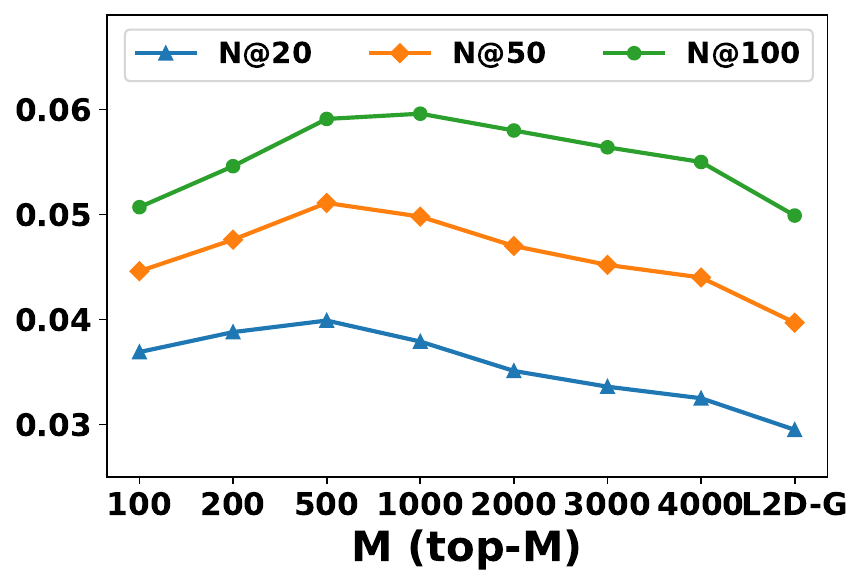}}
        \vspace{-5pt}
  \caption{The impact of $M$ on NDCG metric in the $L2D$-L. $M$ is a hyperparameter that determines the number of hidden states in local aggregation. Note that $L2D$-L equals $L2D$-G when $M$ reaches its maximum length.}
  \label{fig:Mimpact_ndcg}
\end{figure}

In this subsection, we demonstrate the impact of parameter $M$ on the NDCG metric in the $L2D$-L method. As shown in Figure~\ref{fig:Mimpact_ndcg}, the phenomenon observed in the NDCG metric is consistent with the Recall metric in the paper, further strengthening our argument.

\subsection{Compared Methods}
\label{app:Compared Methods.}
In this work, to demonstrate the superiority of our proposed method from the perspective of balancing performance and inference overhead in LLM-based recommendation, we primarily selected some of the most commonly used LLM-based models in the current literature. \wroll{\textbf{For LLM-based embedding}, we included \textbf{AlphaRec} \cite{alpharec} as a baseline. This method uses LLM embeddings for recommendations by applying a collaborative filtering model to utilize language representations. \textbf{For LLM-based generative recommendation}, we included the following methods:
(1)\textbf{BIGRec}~\cite{bigrec}: A generative LLM-based recommender that predicts the next item via historical interactions, mapping generated items to the dataset by L2 distance matching on semantic embeddings. (2) \textbf{GPT4Rec}~\cite{GPT4Rec}: Similar to BIGRec but employs BM25 for item mapping.
(3) $\mathbf{D^3}$~\cite{d3}: 
An improved variant of BIGRec that mitigates decoding bias by eliminating length normalization. Additionally, we included \textbf{non-LLM baselines} (\textbf{SASRec}~\cite{sasrec} and \textbf{GRU4Rec}~\cite{gru4rec}) for comprehensive comparison.}

\clearpage
\newpage

\end{document}